\documentclass[preprint,12pt]{elsarticle}
\usepackage{multirow}
\usepackage{booktabs}  
\usepackage{threeparttable}
\usepackage{amssymb}
\usepackage{caption}
\usepackage{subfigure}
\usepackage{amsmath}
\journal{JINST}
\biboptions{sort&compress}

\begin{document}
\begin{frontmatter}

\title{ Analysis of signal waveform from a midsize liquid argon detector}

\renewcommand{\thefootnote}{\fnsymbol{footnote}}

\author{
K.K.~Zhao$^{a,b}$,
M.Y.~Guan$^{b,c}\footnote{Corresponding author. E-mail address:~dreamy\_guan@mail.ihep.ac.cn. }$,
J.C.~Liu$^{b,c}$,
C.G.~Yang$^{b,c}$,
S.T.~Lin$^{a}$
}

\address{

${^a}${College of Physics, Sichuan University, Chengdu, 610065, China}

${^b}${Institute of High Energy Physics, Chinese Academy of Sciences, BeiJing, 100049, China}

${^c}${University of Chinese Academy of Sciences, BeiJing, 100049, China}
	
}

\begin{abstract}

The midsize single-phase liquid argon prototype detector, operating at the surface laboratory, is designed to measure scintillation light emitted by the liquid argon (LAr). The detector employs 42 8-inch photomultiplier tubes (PMT) to collect the light. By analyzing the waveform of the signal, important detector characteristics such as the slow decay time constant that characterizes the purity of the liquid argon can be obtained. To describe the signal waveform, a model, which takes into account the liquid argon emission decay times together with the TPB re-emission process as well as the signal 
reflection effects, is used. The TPB re-emission process is introduced using a three-exponential time structure. Additionally, experimental results provide comprehensive validation for a post-peak hump structure, which is attributed to signal reflection.

\end{abstract}


\end{frontmatter}


\section{Introduction}\label{sec:section1}

\begin{figure}[htbp]
	\centering
	\includegraphics[width=6cm]{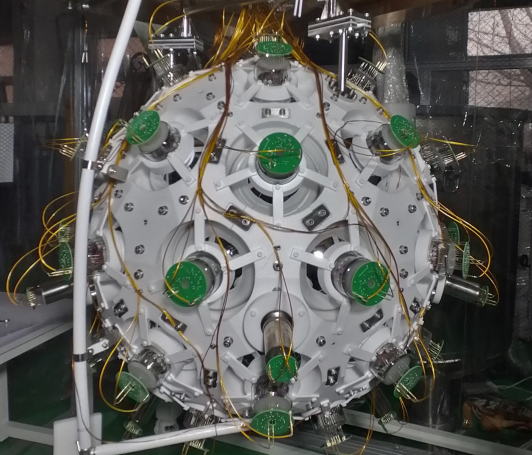}
	\caption{\label{graph_1}The photo of the central detector. The shell in white is PTFE supporting structure. PMTs are placed evenly on the sphere.} 	
\end{figure}

Since January 2022, the ton scale liquid argon prototype detector has been running at the Institute of High Energy Physics of the Chinese Academy of Sciences in Beijing as part of the R\&D project for the future hundred-ton liquid argon dark matter experiment~\cite{1}. The detector comprises 42 8-inch PMTs (Hamamatsu model: R5912-20MOD~\cite{16}) supported by a polytetrafluoroethylene (PTFE) polyhedral spherical structure, as shown in Fig.~$\ref{graph_1}$. The center detector is filled with approximately 2.2 tons of liquid argon, with a volume of about 60 cm in diameter, and the photocathode coverage of the PMTs reaches about 65\%. The detector employs 1,1,4,4-tetraphenyl-1,3-butadiene (TPB)~\cite{17,18} wavelength shifter to convert VUV photons emitted by the liquid argon into 420 nm visible light. The PMT exterior surface is coated with about 160~$\mu$g/cm$^2$ TPB film, and the inner surface of the center detector is coated with 390-490~$\mu$g/cm$^2$ TPB film. To calibrate the energy, a PTFE film-wrapped $^{241}$Am source is placed at the center of the detector. The data acquisition (DAQ) system consists of front-end digital modules (FDM) and a trigger clock module (TCM), designed for 1GSPS/14-bit waveform sampling and clock trigger distribution, respectively. The DAQ system was developed by Shubin Liu’s team at the University of Science and Technology of China~\cite{2}.

The liquid argon signal waveform exhibits a double exponential time structure, which originates from the decay of excited singlet and triplet states corresponding to the fast and slow components~\cite{4,5,6,3,100}. Analysis of the signal waveform depends on whether a wavelength shifter (WLS) is used to detect the liquid argon luminescence. When experiments do not use WLSs, the sum of two exponentials that describe the fast and slow components cannot accurately capture the signal waveform~\cite{7,8}. To address this issue, some researchers introduce a recombination process~\cite{7}. However, early experiments have shown that the recombination time for liquid argon is too short to affect the signal time profile, as evidenced by the comparison of the liquid argon signal waveform obtained with and without an electric field~\cite{9}. In experiments that use WLSs, the signal waveform analysis becomes more complex due to the re-emission process of the WLS. Some authors introduce an intermediate component to the two exponential time structure to fit the liquid argon signal waveforms~\cite{4,10,14}, However, as the authors mentioned that the intermediate component is confirmed by the signal shape fitting. Whether this has a physical origin is still not clear~\cite{4,14}. Others use a non-exponential intermediate component, as well as the response of TPB, detector, and PMT after-pulsing to describe the liquid argon scintillation time profile~\cite{11}. Some researchers attribute the intermediate exponential component to the re-emission process of the WLS and propose a complex TPB re-emission model~\cite{12}. However, this model is not consistent with our detector data after testing due to the presence of microsecond-scale time structure.

Several authors have reported the presence of a hump structure after the peak of the liquid argon signal waveform~\cite{10,11,13}, based on data acquisition using signal waveform digitization electronics. In Ref~\cite{10}, the authors suggest that signal reflections at the flange feed-through occur approximately 200 ns and 400 ns after the maximum, but without further explanation. Ref~\cite{11} indicates that a mismatch of around 9\% between the data and signal model occurs at approximately 100 ns. Ref~\cite{13} reports the observation of a hump structure at approximately 60 ns, with no conclusive explanation.

Similarly, in our experimental data, we also observed a hump structure. To investigate if the hump is caused by signal reflection, we conducted two tests. The first test involved comparing the hump position changes obtained from PMTs with different cable lengths. The second test was to measure the PMT in-situ response on a picosecond pulsed light source, looking for signal signatures corresponding to the time of the hump. Through the analysis of the ton scale liquid argon prototype detector signal data and a comparison with the literature, we propose a model that can fairly fit the signal waveform data.

\section{Data analysis and comparison}\label{sec:section2}

\subsection{Scintillation signal model derivation}
The average signal waveform of liquid argon V(t) usually can be simply written as follows~\cite{14}:
\begin{equation}
V(t)=S(t) \otimes R(t)
\end{equation}
where S(t) is the scintillation signal of the combination of liquid argon and the TPB. where R(t) is the response function of the signal detection and data acquisition system (PMTs and DAQ). This part can be represented by a Gaussion function with a standard deviation $\sigma$.

The liquid argon scintillation decay process is characterized by a double-exponential time structure, which corresponds to two excited states~\cite{3}. Recent studies suggest that TPB exhibits delayed light emission~\cite{12}. To account for this phenomenon, a three-exponential TPB reemission response is proposed based on mathematical decay time theory~\cite{15}. The scintillation signal equation S(t) is obtained by convolving the sum of liquid argon scintillation with the TPB response. The intermediate component is believed to be associated with the luminescence process of TPB, as reported in Ref~\cite{12}.

\begin{footnotesize}
\begin{equation}
S(t)=\left\{\frac{A_s}{\tau_{\mathrm{s}}} \exp \left(-\frac{t}{\tau_s}\right)+\frac{A_T}{\tau_T} \exp \left(-\frac{t}{\tau_T}\right)\right\} \otimes \sum_{j=1, 2, 3} \frac{R_j}{\tau_j} \exp \left(-\frac{t}{\tau_j}\right)
\end{equation}
\end{footnotesize}
where $A_S$ and $A_T$ respectively represent the intensity of the fast and slow components of liquid argon scintillation. $\tau_{\mathrm{S}}$ and $\tau_{\mathrm{T}}$ respectively represent the time constants of the fast and slow components. $R_1$, $R_2$ and $R_3$ are the proportions of the three re-emission components of TPB ($R_1$+$R_2$+$R_3$=1). $\tau_1, \tau_2$ and $\tau_3$ are the three re-emission time constants of TPB.

\subsection{Explanation of the hump}
\subsubsection{Hump position changes with cable length}

\begin{figure}[htbp]
\centering    
 
\subfigure[] 
{
	\begin{minipage}{6cm}
	\centering          
	\includegraphics[scale=0.35]{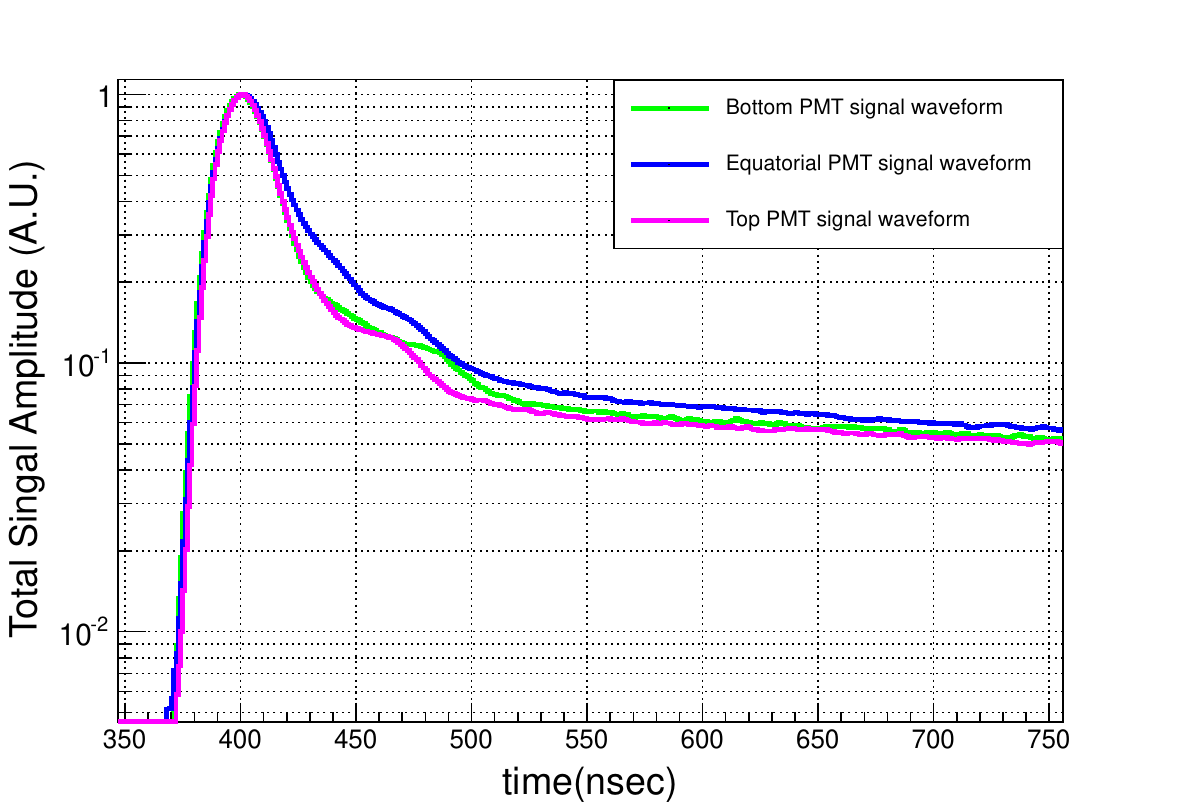}   
	\end{minipage}
}
	 \qquad
\subfigure[] 
{
	\begin{minipage}{6cm}
	\centering      
	\includegraphics[scale=0.31]{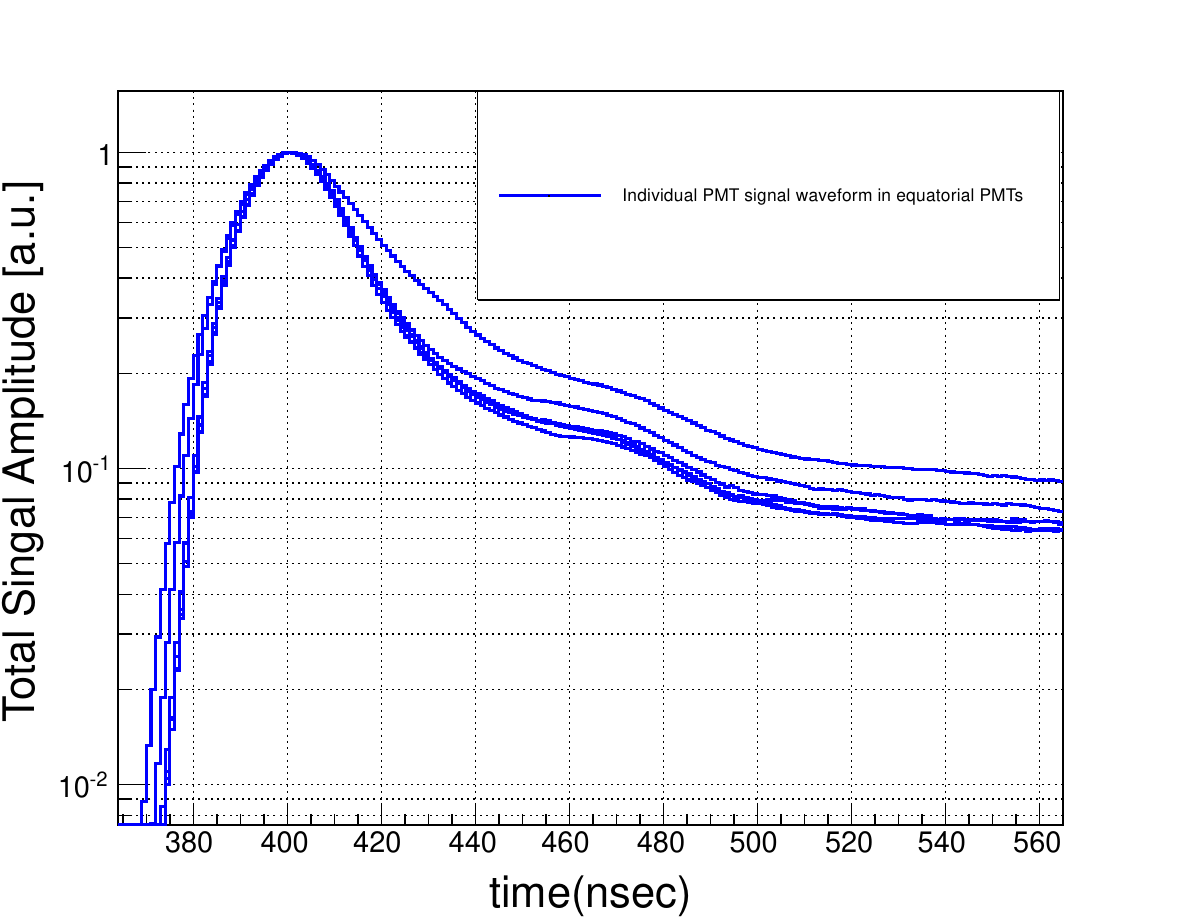}   
	\end{minipage}
}
 
  \caption{\label{graph_3} (a) The averaged gamma-rays signal waveforms of the same run coming from the top PMT, the equatorial PMT and the bottom PMT of the central detector. The corresponding cable lengths between the feedthrough and the PMT base are 1.9~m, 2.5~m and 3.5~m, respectively. (b) These blue lines represent signal waveforms from six PMTs locating at the equator, respectively. }
\end{figure}

In the ton scale liquid argon detector, the cable length between the PMTs and the feedthrough is not uniform. As the center detector extends from top to bottom, the cable length gradually increases. However, for PMTs on the same horizontal circle, the cable length is equal. The maximum difference in cable length is 1.6m between the top and bottom PMTs. Fig.~$\ref{graph_3}$ displays the signal waveforms for the top PMT, six equatorial PMTs, and the bottom PMT.

From Fig.~$\ref{graph_3}$a, it is evident that the hump positions differ based on cable length. For the six equatorial PMTs with identical cable length, the hump positions are nearly identical, as seen in Fig.~$\ref{graph_3}$b. The top PMT, with the shortest cable length between the feedthrough and PMT base, shows the hump first, followed by the equatorial PMT and the bottom PMT. The time differences between the three humps correspond to the length differences of the three cables.

\begin{figure}[htbp]
\centering    
 
\subfigure[] 
{
	\begin{minipage}{6cm}
	\centering          
	\includegraphics[scale=0.35]{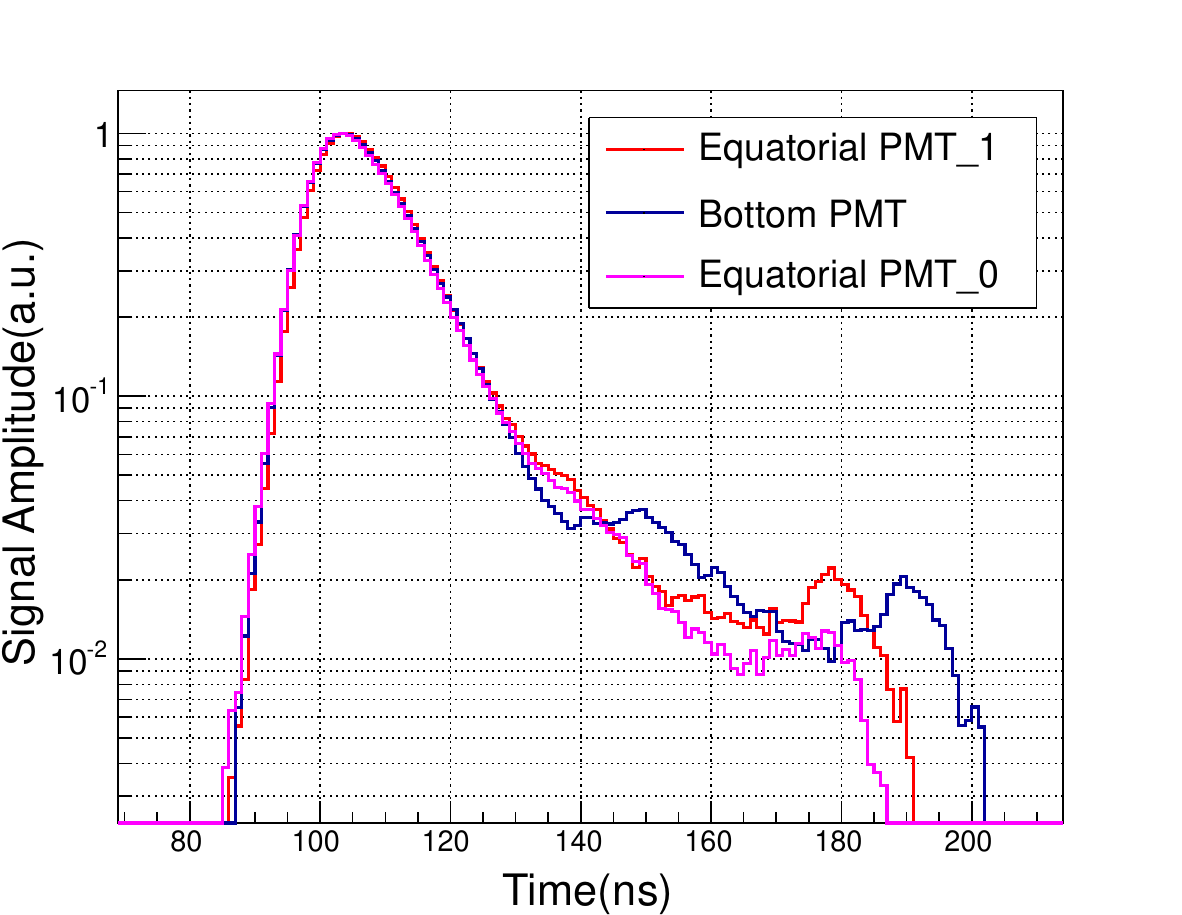}   
	\end{minipage}
}
	 \qquad
\subfigure[] 
{
	\begin{minipage}{6cm}
	\centering      
	\includegraphics[scale=0.35]{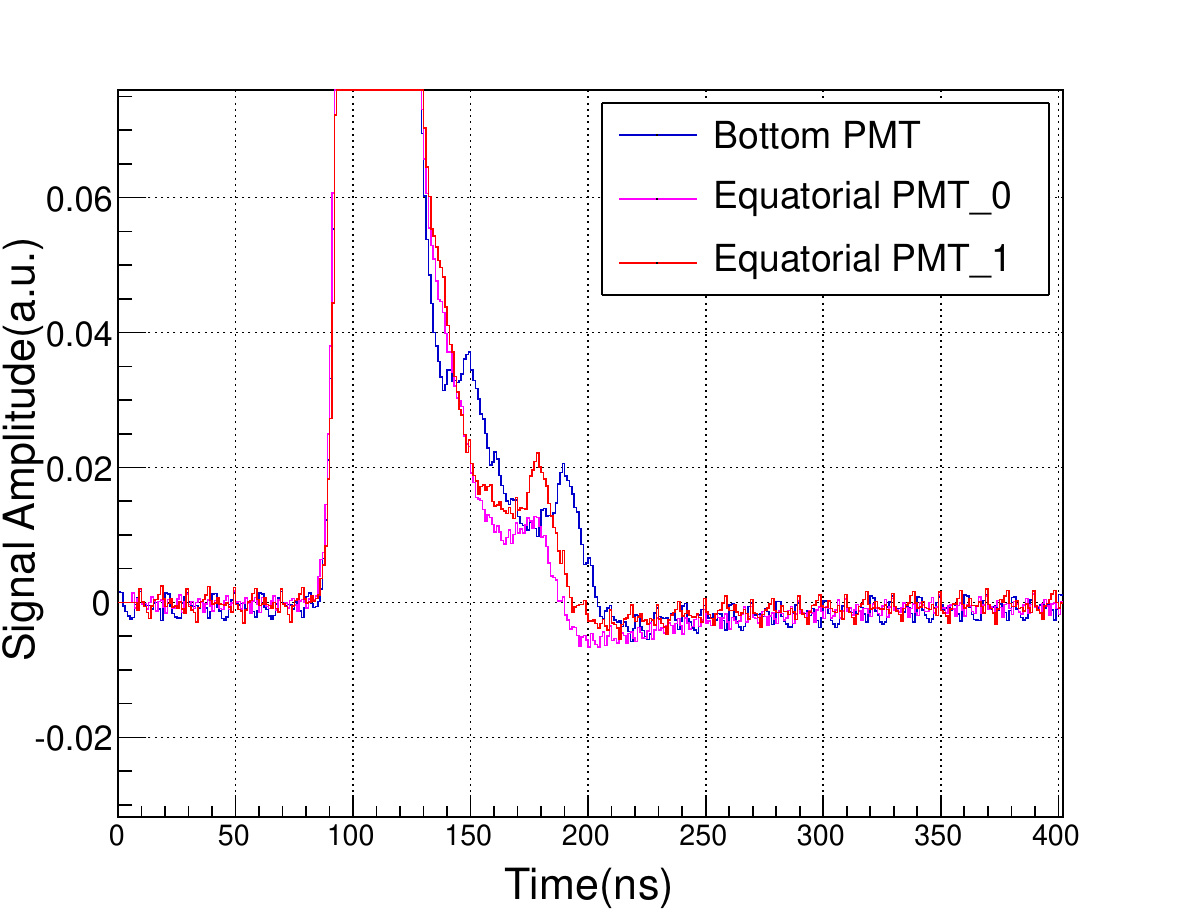}   
	\end{minipage}
}
 
  \caption{\label{ref} The averaged signal waveforms of the bottom PMT and two equatorial PMTs triggered by the picosecond pulsed laser. (a) Three signal waveforms with two different cable lengths. (b) The details of the signal waveform expressed in linear coordinates. }
\end{figure}

\subsubsection{Piosecond pulsed laser experiment}

An experiment using a picosecond pulsed laser was conducted to further investigate if signal reflection occurred in the PMT cable. A PicoQuant model LDH-PC-405 pulsed laser head was utilized as a light source to illuminate the detector PMTs through a fiber optic feedthrough, producing a 405 nm light pulse with a 100 ps pulse width (FWHM). According to Ref~\cite{101}, the absorption re-emission rate of TPB at 405 nm is very low and will not affect the pulsed
laser experiment. The detector DAQ system was connected to a synchronizing trigger signal from the PicoQuant model PDL-800D laser driver. The R5912-20MOD PMT has a typical TTS (Transit Time Spread) of 3 ns according to the datasheet. Thus, the signal produced by the laser pulse on the PMTs is equivalent to the signal of many photoelectrons simultaneously superimposed. Fig.~$\ref{ref}$ depicts the signal results of two equatorial PMTs and a bottom PMT. The falling edge of the signal caused by the picosecond pulsed laser is much faster than the liquid
argon signal in Fig.~$\ref{graph_3}$. The signal results show two sets of obvious reflections
on all three signals. The location of the reflection signals varies with the length of the cables. It can be observed that the second reflection peak in Fig.~$\ref{ref}$a is almost at the same timing as the humps in Fig.~$\ref{graph_3}$. The first reflection signal peaks of the equatorial PMTs are likely to be partly overpowered by the falling edge of the signal in Fig.~$\ref{ref}$a due to their shorter cable length. Therefore, it is reasonable to deduce that the possible humps in Fig.~$\ref{graph_3}$ corresponding to the first reflection in Fig.~$\ref{ref}$a are likely to be overpowered by the much slower falling edge of the liquid argon signal.

\subsubsection{Reflection model}

\begin{figure}[htbp]
\centering    
 
\subfigure[] 
{
	\begin{minipage}{8cm}
	\centering          
	\includegraphics[scale=0.25]{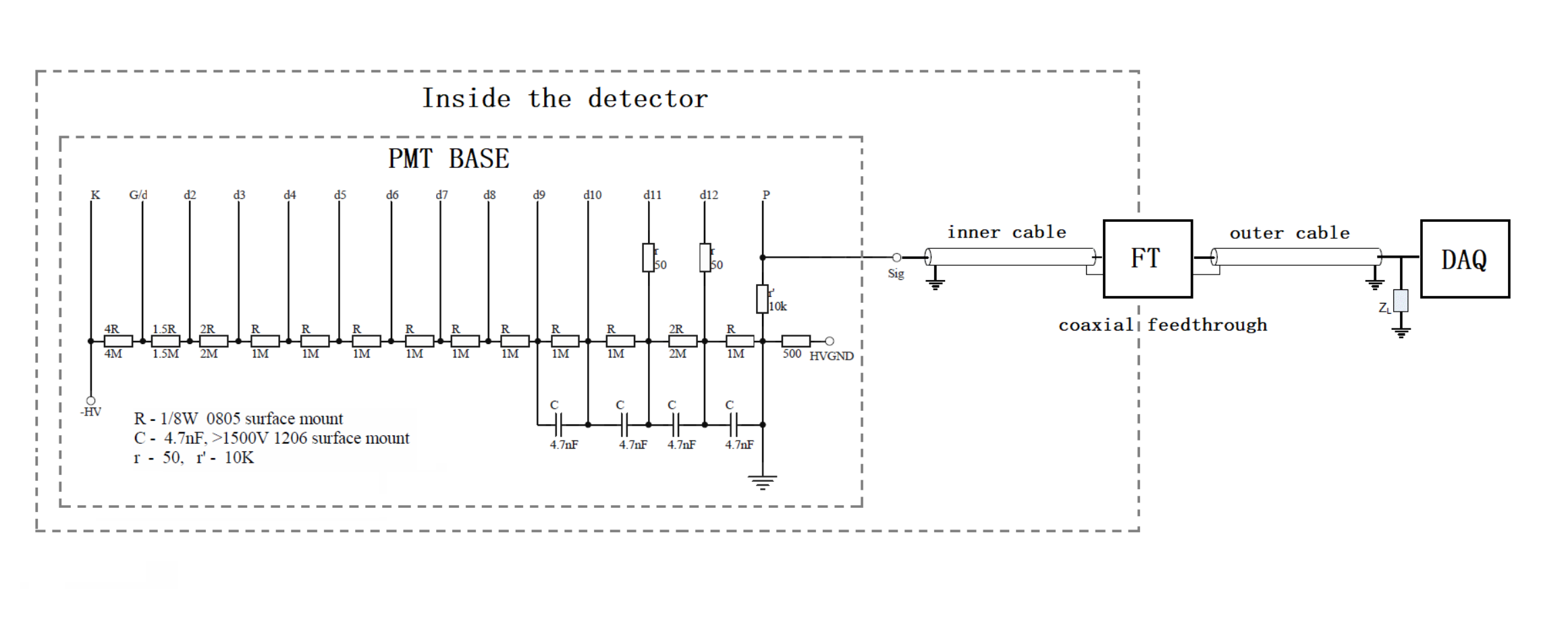}   
	\end{minipage}
}
	 \qquad
\subfigure[] 
{
	\begin{minipage}{4cm}
	\centering      
	\includegraphics[scale=0.075]{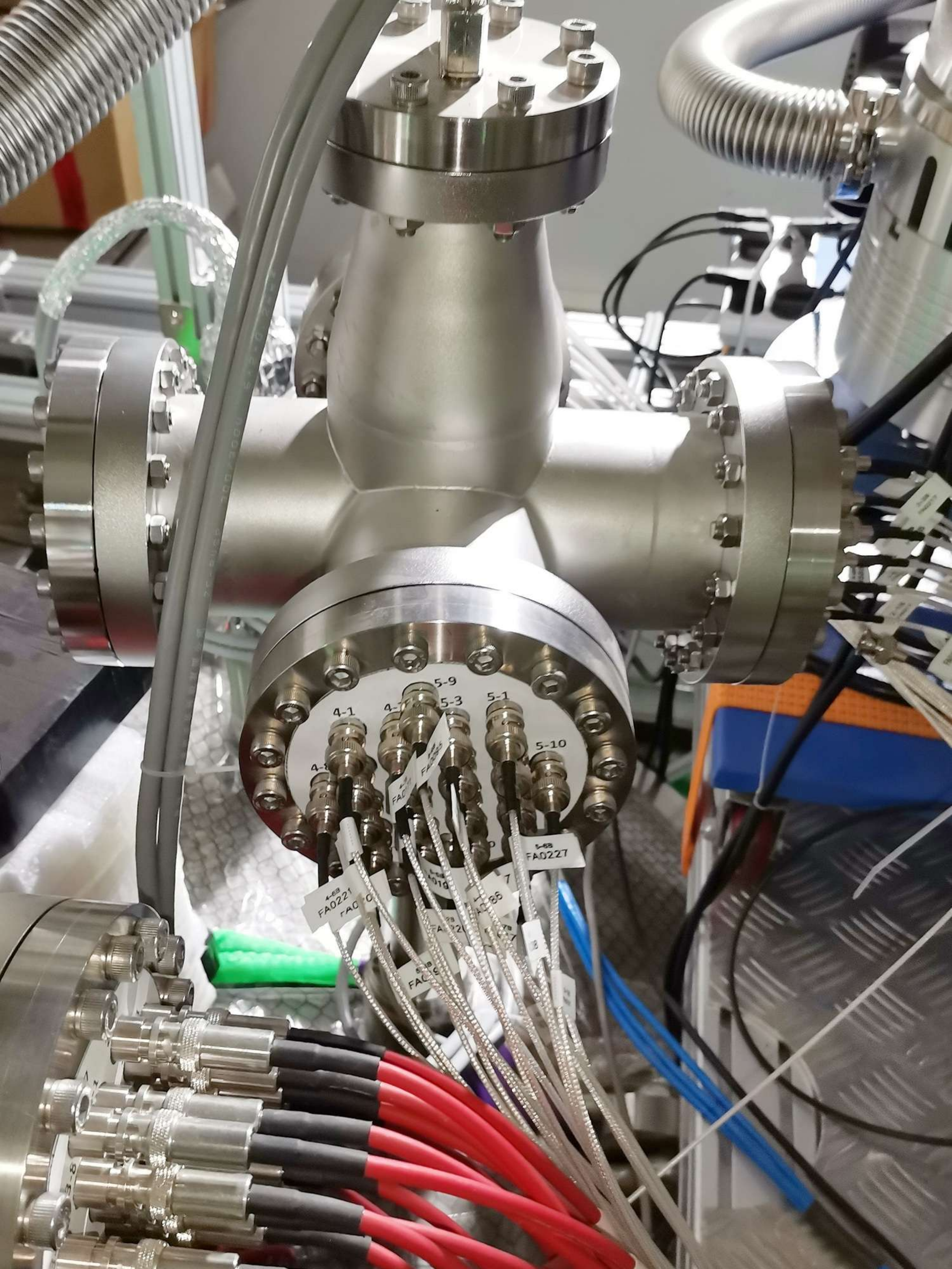}   
	\end{minipage}
}
 
  \caption{\label{daq}(a) Electrical scheme of the PMT voltage divider and the signal cable connection of the detector.(b) Coaxial cable feedthrough.}
\end{figure}

Fig.~$\ref{daq}$ illustrates the electrical arrangement of the PMT voltage divider and signal cable connection. The inner cable is a PI-insulated coaxial cable with a 50~$\Omega$ impedance and a length of approximately 1.9-3.5 meters. The outer cable is an RG 316 coaxial cable with a length of 4.5 meters. The signal feedthrough is a BNC coaxial feedthrough with 20 pins and a grounded shield from MPF Products, Inc. It should be noted that the nominal impedance of the cables and the feedthrough is 50~$\Omega$. The signal cable is directly connected to the PMT anode to increase the signal output. Far-end parallel termination (ZL = 50~$\Omega$) is employed to eliminate signal reflection. Based on the measurements presented in Fig.~$\ref{ref}$, it can be inferred that a signal reflectance of approximately a few percent exists along the inner cable. If the cable length is not long enough, the falling edge of the liquid argon signal may overpower the first reflection peak. A relatively small undershoot  at about 200ns shows up in Fig.~$\ref{ref}$b which expressed the details of the signal waveform in linear coordinates. The results of the picosecond pulsed laser experiment provide strong evidence for the reflection cause of the hump. Only the second reflection signal appears clearly in the LAr signal waveform. A simple model suggests that the hump is formed due to two reflections of the signal at the feedthrough and the PMT base, which are caused by small impedance mismatch at the feedthrough.

\begin{equation}
I(t)=V\left(t-t_0\right)+A^* V\left(t-\left(t_0+2^*\triangle t\right)\right)+B^* V\left(t-\left(t_0+4^*\triangle t\right)\right)
\end{equation}

where $t_0$ is the scintillation time of liquid argon. $\triangle t$ is the time delay of the inner cable. A and B correspond to the intensity of  reflection respectively. The factor 2 and 4 is according to the reflection sequence.

Fig.~$\ref{graph_4}$ displays the fitting outcomes of signal data from 10 PMTs at the detector equator, using the same cable length for $^{241}$Am 59.5 keV gamma events. The results show a relatively small difference between the experimental data and equation (3), indicating that the model accurately describes the signal waveform's details. The fitting parameters are presented in Table~$\ref{tab1}$. The fitting time range is selected according to 3~$\tau_T$ in order to account for the hump feature. This time interval enables a good description of LAr's slow component decay process. The parameter values listed in Table~$\ref{tab1}$ indicate that: 
\begin{figure}[htbp]
	\centering
	\includegraphics[width=13cm]{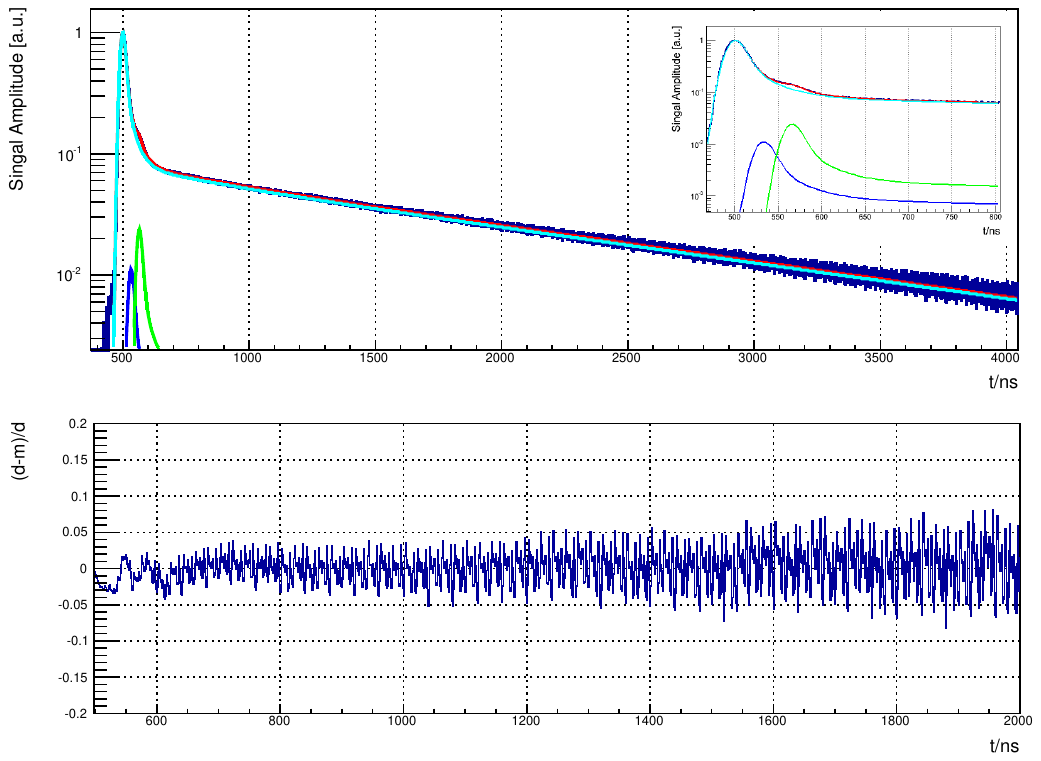}
	\caption{\label{graph_4} The averaged signal waveform of 10 PMTs at the detector equator with the same cable length for the $^{241}$Am 59.5~keV gamma events. A fit to the data is shown, using Eq. (3). The colors in the figure are as follows: Light blue, blue and green represent the three items of equation (3) from left to right. The inset shows the fitting results near the hump in the corresponding waveform. The fitting time range is chosen according to 3$\tau_{\mathrm{T}}$.} 	
\end{figure}

\begin{table}[htbp]
    \footnotesize
	\centering
	\caption{Fit parameters(using Eq. (3)). These parameters are for the equator PMTs. }
	\label{tab1}  
	\begin{tabular}{ccccccccccc} 
    \toprule 
    \multicolumn{3}{c}{}&\multicolumn{3}{c}{LAr and reflection}&&\multicolumn{3}{c}{TPB}\\
    \multicolumn{3}{c}{}&Par&gamma-rays&&
                        &Par&gamma-rays&&\\      
    \hline 
    \multicolumn{3}{c}{}&$A_S$&43.74$\pm$1.48&& &$R_1$&0.23$\pm$0.27&\\      
    \multicolumn{3}{c}{}&$A_T$&92.84$\pm$1.83&& &$R_3$&0.24$\pm$0.15&\\
    \multicolumn{3}{c}{}&$\tau_{\mathrm{S}}$(ns)&6.5$\pm$5.5&& &$\tau_1$(ns)&1.7$\pm$0.001&\\
    \multicolumn{3}{c}{}&$\tau_{\mathrm{T}}$(ns)&1482$\pm$1.4&& &$\tau_2$(ns)&31.8$\pm$1.60&\\
    \multicolumn{3}{c}{}&A&0.011$\pm$0.223&& &$\tau_3$(ns)&329$\pm$1.4&\\
    \multicolumn{3}{c}{}&B&0.024$\pm$0.078&\\
    \multicolumn{3}{c}{}&$t_0$(ns)&492.5$\pm$0.1&\\
    \multicolumn{3}{c}{}&$\triangle t$(ns)&16.32$\pm$1.19&\\
    \multicolumn{3}{c}{}&$\sigma$&9.41$\pm$0.007&& &~~~~~~~~~~~~~*~$R_1$+$R_2$+$R_3$=1\\
    \bottomrule 
    \end{tabular}
\end{table}

1. The inner cable time delay estimated from the fit is approximately 16.3 ns, which is reasonable for the 2.5 m cable length.
2. The intensity of the reflection is about 1.1\% and 2.4\% according to the time of 2$\triangle t$ and 4$\triangle t$. Since there is no corresponding peak on the signal waveform at 2$\triangle t$, parameter degeneracy of this reflection peak is inevitable, which may result in A fitting smaller than B. 
3. $t_0$ is about 7.5 ns ahead of the waveform peak, which is caused by the TPB light absorption reemission process.
4. The fast decay constant obtained from the fit is about 6.5ns, which is consistent with literatures.
5. The first time constant of TPB obtained from the fit is much smaller than liquid argon's fast time constant. The value of the fast decay time is similar to data reported for another WLS of POPOP~\cite{4}.
6. The slow component decay time constant estimated from the fit is 1482 ns, which aligns with the value reported in Ref~\cite{10,11}.

\section{Conclusions}\label{sec:section3}

Nowadays, waveform digitization electronics are widely utilized for processing liquid argon signals. The signal waveform contains crucial information on signal generation and transmission. It should be noted that for liquid argon detectors, the internal cables and the external cables are connected via feedthroughs. Part of the internal cable is kept in cryogenic temperature and may not be of the same model as the external cable. This configuration may cause minor impedance mismatch problems. This paper aims to explain and validate the formation of a hump structure that occurs after the signal peak. Experimental data reveals that the hump is caused by signal reflections, and its position is affected by signal reflections between the feedthrough and the PMT base. This indicates that signal transmission can induce changes to the signal waveform. Nonetheless, this paper introduces a method for in-situ measurement of signal reflection using a picosecond pulsed laser. After incorporating the reflection effect into the signal waveform model, it fits well with the experimental data.

Additionally, this work describes a model for interpreting liquid argon signal generation. In this model, the emission decay times of liquid argon is incorporated with the response of TPB. Based on the delayed component of TPB re-emission, a three-exponential time structure of TPB re-emission is introduced. Full data fitting results indicate that the 1st time constant of TPB is approximately 1.7 ns, which is significantly shorter than liquid argon's fast time constant. The 2nd time constant of TPB is about 31.8 ns, which is similar to the published literature~\cite{10,12}. The 3rd time constant is approximately 329 ns. Data analysis demonstrates that the slow decay time constant of liquid argon scintillation light is roughly 1482 ns, which is similar to the value reported in Ref~\cite{10,11}.

By comparing with the fit results of the model using three decay times~\cite{4,10,14} (appendix), we found that the difference in the slow component time constant obtained by two LAr signal models is about 3.4\%. Similar fast component time constants can be obtained by two LAr signal models. These small difference shows that the new waveform model can not only give right liquid argon slow component time constant, but also describe the hump structure in signal waveform well.

\section{Acknowledgments}

The study is supported by the National Key Research and Development Program of the People’s Republic of China (2016YFA0400304). The authors would like to thank Qin Zhao, Zhimin Wang and Lei Wang for the helpful discussion.

\section{Appendix  Fitting results using three decay times}

The pulse shape model of considering three decay times of liquid argon can be expressed by the following function~\cite{4,10,14}: 

\begin{small}
\begin{equation}
\begin{aligned}
f(t)= \sum_{j=\text { fast,slow,int }} \frac{2 A_j}{\tau_j} \exp \left[\frac{\sigma^2}{2 \tau_j^2}-\frac{t-t_0}{\tau_j}\right]
\times\left(1-\operatorname{Erf}\left[\frac{\sigma^2-\tau_j\left(t-t_0\right)}{\sqrt{2} \sigma \tau_j}\right]\right)
\end{aligned}
\end{equation}
\end{small}

\begin{figure}[htbp]
	\centering
	\includegraphics[width=13cm]{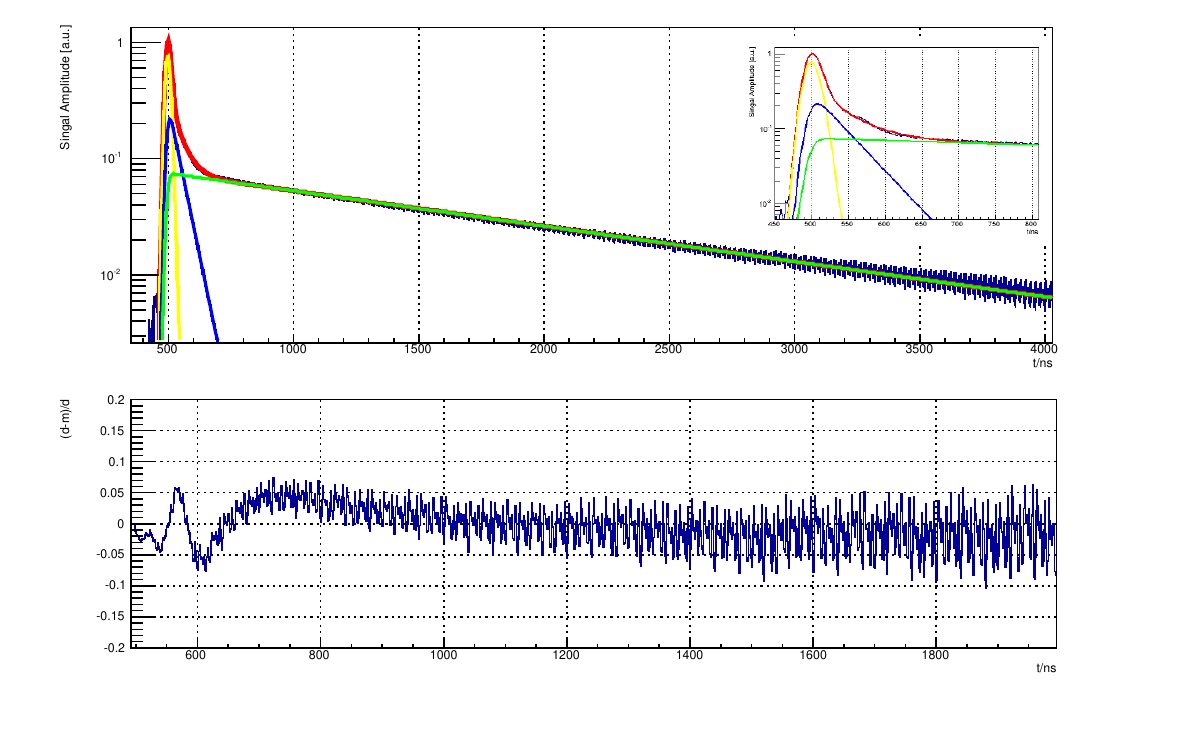}
	\caption{\label{graph_5}The fitting results using equation (4). the signal waveform data are the same as Fig.~$\ref{graph_4}$. The colors in the figure are as follows: yellow, blue and green represent the fast, intermediate and slow components of the liquid argon decay times.} 	
\end{figure}

\begin{table}[htbp]
    \footnotesize
	\centering
	\caption{Fit parameters(using Eq. (4)). }
	\label{tab2}  
	\begin{tabular}{ccccccccccc} 
    \toprule 
  
    \multicolumn{3}{c}{}&Par&gamma-rays&&
                        &Par&gamma-rays&&\\      
    \hline 
    \multicolumn{3}{c}{}&$A_S$&5.31$\pm$3.95&& &$\tau_{\mathrm{S}}$(ns)&6.5$\pm$21.6\\    
    \multicolumn{3}{c}{}&$A_I$&3.37$\pm$3.63&& &$\tau_{\mathrm{I}}$(ns)&42.7$\pm$62.6&\\ 
    \multicolumn{3}{c}{}&$A_T$&26.54$\pm$2.88&& &$\tau_{\mathrm{T}}$(ns)&1431$\pm$235&\\
    \multicolumn{3}{c}{}&$\sigma$&9.46$\pm$4.8&& \\
    \multicolumn{3}{c}{}&$t_0$&494.2$\pm$13.1&& \\
    \bottomrule 
    \end{tabular}
\end{table}

where the response function of the signal detection and data acquisition system (PMTs and DAQ) is included accordingly. The fitting results are shown in Fig.~$\ref{graph_5}$ and Table~$\ref{tab2}$. The results show a relatively bigger difference between the experimental data and equation (4) at 550ns-800ns, indicating that the model can not resolve the "hump" problem. As a comparison, the slow component decay time constant obtained by equation (4) differs by about 3.4\% from that obtained by equation (3).

\end{document}